\documentclass[12pt]{article}
\usepackage{graphicx}
\def \app{D_{\pi \pi}}

\def \bbpp{\overline{{\cal B}}_{\pi \pi}}
\def \bea{\begin{eqnarray}}
\def \beq{\begin{equation}}
\def \bo{B^0}
\def \ko{K^0}

\def \cn{Collaboration}
\def \cpp{C_{\pi \pi}}
\def \eea{\end{eqnarray}}
\def \eeq{\end{equation}}
\def \ite{{\it et al.}}

\def \lpp{\lambda_{\pi \pi}}
\def \ob{\overline{B}^0}
\def \ok{\overline{K}^0}

\def \spp{S_{\pi \pi}}
\begin{document}

\begin{flushright}
TECHNION-PH-2002-21 \\
EFI 02-81 \\
hep-ph/0205323 \\
May 2002 \\
\end{flushright}

\renewcommand{\thesection}{\Roman{section}}
\renewcommand{\thetable}{\Roman{table}}
\centerline{\bf CONVENTION-INDEPENDENT STUDY OF CP-VIOLATING}
\centerline{\bf ASYMMETRIES IN $B \to \pi \pi$}
\medskip
\centerline{Michael Gronau}
\centerline{\it Physics Department, Technion -- Israel Institute of Technology}
\centerline{\it 32000 Haifa, Israel}
\medskip
\centerline{Jonathan L. Rosner}
\centerline{\it Enrico Fermi Institute and Department of Physics}
\centerline{\it University of Chicago, Chicago, Illinois 60637}
\bigskip

\begin{quote}

CP-violating asymmetries in the decay $B^0(t)\to \pi^+ \pi^-$ are a potentially
rich source of information about both strong and weak phases.  In a previous
treatment by the present authors use was made of an assumption about the
relative magnitude of tree and penguin amplitudes contributing to this
process.  This assumption involved an ambiguity in relating the tree amplitude
to the amplitude for $B \to \pi \ell \nu$.  It is shown here that one can avoid
this assumption, which adopted a particular convention for tree and penguin
amplitudes, and that the results are convention-independent.

\end{quote}

\leftline{\qquad PACS codes:  12.15.Hh, 12.15.Ji, 13.25.Hw, 14.40.Nd}

\section{Introduction}

The study of CP-violating asymmetries in the decays $B^0(t) \to \pi^+ \pi^-$
has reached an interesting stage.  Two collaborations working at asymmetric
$B$ factories, the Babar Collaboration at PEP-II (Stanford) \cite{CSBa} and 
the Belle Collaboration at KEK-B (Tsukuba, Japan) \cite{bCSBe} have both 
reported measurements of time-dependent asymmetries in this process and its
charge-conjugate which are potentially rich sources of information of both
strong and weak phases.  The weak phases are those of elements in the
Cabibbo-Kobayashi-Maskawa (CKM) matrix describing the weak charge-changing
couplings of quarks.  At present these phases provide a satisfactory
description of all observed CP-violating phenomena in both $K$ and $B$ decays.

In a previous article \cite{GR02} (for a more complete discussion, see also 
\cite{GR01}), we analyzed these CP-violating asymmetries using assumptions
which included knowledge of the ratio of tree and penguin amplitudes 
\cite{PT,FlMat}. This knowledge was obtained from other processes using the
factorization hypothesis.  However, the nature of the tree amplitude 
and the value of the above ratio depended on our convention for defining the 
tree and penguin amplitudes, leading to some indeterminacy in the result.
Certain aspects of ambiguities following from the penguin amplitude convention
were discussed earlier in \cite{BF,CP,LSS}, and recently in \cite{DL}. 

In the present paper we find that one can obtain useful information from
CP-violating asymmetries in $B^0 \to \pi^+ \pi^-$ {\it independently} of
the penguin amplitude convention, and without prior knowledge of the
tree/penguin ratio.  Some sacrifice in statistical power unavoidably occurs,
so that determination of the weak phase $\alpha = \phi_2$ to better than
$10^\circ$ is difficult without additional assumptions.  Thus, $\Delta
\alpha \simeq 10^\circ$ seems to be an estimate of the theoretical systematic
error of the present method.  This would still represent an improvement with
respect to the present situation, in which we estimated $\alpha$ to be
determined only within a $50^\circ$ range \cite{GR02}.

The data which we use in the present determination consist of the
charge-averaged branching ratio $\bbpp$, the time-dependent asymmetries
$\spp$ and $\cpp$ which are coefficients of $\sin \Delta m t$ and
$\cos \Delta m t$, and the charge-averaged branching ratio ${\cal B}(B^\pm
\to K \pi^\pm)$.  Similar inputs were also advocated in a previous analysis
by Charles \cite{JC}, which differs in details of correction factors and
which presents results in terms of the $\rho$ and $\eta$ variables of the 
CKM matrix \cite{WP} rather than in terms of the phase $\alpha$.

The paper is organized as follows.  We introduce two different amplitude
conventions in Section II.  We show that, while the tree amplitudes in the two
conventions are different, the corresponding penguin amplitudes are essentially
the same, up to a simple CKM factor.  We write down a dictionary relating 
the magnitudes and strong phases of corresponding tree amplitudes.  In Section
III we specify our assumptions and explain the method for determining 
the weak phases $\gamma$ or $\alpha$, as well as the relevant strong phase, 
by including information about the penguin amplitude in $B^+ \to K^0 \pi^+$.
The only required assumptions are penguin dominance of this amplitude and 
factorization of penguin amplitudes.  We also summarize the present relevant
experimental data.  In Section IV we then plot the two measured CP-violating
asymmetries as functions of strong and weak phases. We also plot
relations between strong phases in the two conventions.
While no use is made in this study of a prior knowledge of the ratio of
tree and penguin amplitudes, this ratio could be used as a cross check and 
could resolve a possible discrete ambiguity in determining the weak phase.
Section V qualitatively compares uncertainties in evaluating this ratio in 
the two conventions using other experimental inputs.
Experimental prospects and conclusions are contained in Section VI.

\section{Notations and conventions}

The expressions for the decay amplitudes of $\bo \to \pi^+ \pi^-$ and $\ob 
\to \pi^+\pi^-$ depend on the convention employed. We now describe two 
different conventions used in the literature, denoted $c$ and $t$ conventions, 
where $c$ and $t$ represent appropriate CKM factors governing penguin
amplitudes.
\bigskip

\leftline{\bf A.  $c$ convention}
\medskip

In the convention of Refs.\ \cite{GR02,GR01}, one writes the decay amplitudes
in terms of a color-favored tree amplitude $T_c$ and a penguin amplitude $P_c$
as
$$
A(\bo \to \pi^+ \pi^-) = -(|T_c|e^{i \delta^T_c} e^{i \gamma} +
 |P_c| e^{i \delta^P_c})~~~,
$$
\beq \label{eqn:Bppc}
A(\ob \to \pi^+ \pi^-) = -(|T_c|e^{i \delta^T_c} e^{- i \gamma} +
 |P_c| e^{i \delta^P_c})~~~,
\eeq
where we use the definitions in \cite{PDG} of weak phases $\alpha = \phi_2$,
$\beta = \phi_1$, and $\gamma = \phi_3$.  The strong phases of the tree and
penguin amplitudes are $\delta^T_c$ and $\delta^P_c$, while $\delta_c \equiv
\delta^P_c - \delta^T_c$.  Here the subscript $c$ refers to the convention
in which the weak phase of the strangeness-preserving ($\Delta S = 0$) penguin
amplitude in $\bar b \to \bar d q \bar q$ is defined to be that of $V^*_{cb}
V_{cd}$.  The top quark in the $\bar b \to \bar d$ loop diagram has been
integrated out and the unitarity relation $V^*_{tb}V_{td} = -V^*_{cb}V_{cd}
-V^*_{ub}V_{ud}$ has been employed.  The term $-V^*_{ub}V_{ud}$ has been
included in the tree amplitude, which has the same weak phase.
\bigskip

\leftline{\bf B.  $t$ convention}
\medskip

A different convention has been commonly employed in the past \cite {SW} and
also quite recently \cite{LRpipi}.  In this convention, one uses the unitarity
relation in the form $V^*_{cb}V_{cd} = -V^*_{tb}V_{td} -V^*_{ub}V_{ud}$ and
assumes the penguin amplitude to be dominated by the $t$ quark term
$V^*_{tb}V_{td}$.  The tree amplitude, again, absorbs a penguin contribution
proportional to $V^*_{ub}V_{ud}$, but it is different than that in the previous
convention.  For this convention we shall use a subscript $t$ on all
quantities.  The expressions for the decay amplitudes are then
$$
A(\bo \to \pi^+ \pi^-) = -(|T_t|e^{i \delta^T_t} e^{i \gamma} +
 |P_t| e^{i \delta^P_t} e^{- i \beta})~~~,
$$
\beq \label{eqn:Bppt}
A(\ob \to \pi^+ \pi^-) = -(|T_t|e^{i \delta^T_t} e^{- i \gamma} +
 |P_t| e^{i \delta^P_t} e^{i \beta})~~~,
\eeq
where one denotes $\delta_t \equiv \delta^P_t - \delta^T_t$. 
\bigskip

\leftline{\bf C.  Equivalence of the two conventions}
\medskip

It is obvious that the $c$ and $t$ conventions are equivalent. However, since
in general they imply different tree and penguin amplitudes, an assumption
about the tree amplitude in one convention is not equivalent to the same
assumption in the other convention.  On the other hand, as we will show now,
the penguin amplitudes in the two conventions are equal, up to a trivial CKM
factor.  Let us write the amplitude for $B^0\to\pi^+\pi^-$ in a most general
form in terms of the three CKM factors and corresponding three hadronic weak
amplitudes $A_i~(i = u, c, t)$ involving strong phases:
\beq
A(B^0\to \pi^+\pi^-) = V^*_{ub}V_{ud} A_u +  V^*_{cb}V_{cd} A_c
+  V^*_{tb}V_{td} A_t~~.
\eeq
Using unitarity, this can be written in the $c$ and $t$ conventions
as
\bea\label{c}
A(B^0 \to \pi^+\pi^-) &=& V^*_{ub}V_{ud} (A_u - A_t) +  V^*_{cb}V_{cd} (A_c
-A_t)\\
\label{t}
&=& V^*_{ub}V_{ud} (A_u - A_c) +  V^*_{tb}V_{td} (A_t
-A_c)~~.
\eea

Comparing the second terms in Eqs.~(\ref{eqn:Bppc}) and (\ref{eqn:Bppt})
with the corresponding terms in Eqs.~(\ref{c}) and
(\ref{t}), one finds a simple relation between the two penguin amplitudes:
\beq\label{PtPc}
\frac{|P_t|}{|P_c|} = \frac{| V^*_{tb}V_{td}|}{|V^*_{cb}V_{cd}|} = 
\frac{\sin\gamma}{\sin\alpha}~~,~~~~\delta^P_t = \delta^P_c~~.
\eeq
Namely, the penguin amplitudes in the two conventions involve a common 
hadronic matrix element $A_t - A_c$ but different CKM factors.

On the other hand, the relation between tree amplitudes in the
two conventions is more complicated. It can be obtained by subtracting the 
first terms in Eqs.~(\ref{eqn:Bppc}) and (\ref{eqn:Bppt}) from each other 
and comparing with Eq.~(\ref{c}) or (\ref{t}), in which the corresponding 
difference is proportional to the penguin amplitudes, $A_t - A_c$,
\beq
|T_t|e^{-i\delta_t} - |T_c|e^{-i\delta_c} = \frac{|V^*_{ub}V_{ud}|}
{|V^*_{tb}V_{td}|} |P_t| = \frac{\sin\beta}{\sin\gamma} |P_t|
= \frac{\sin\beta}{\sin\alpha} |P_c|~~.
\eeq 
As a consequence of these relations, one has a ``dictionary'' relating
the two conventions, with
\beq
|P_t| \sin \alpha = |P_c| \sin \gamma~~,~~~
|T_t| \sin \delta_t = |T_c| \sin \delta_c~~~,
\eeq
\beq
X_t \cos \delta_t \sin \gamma - X_c \cos \delta_c \sin \alpha = \sin \beta~~~,
\eeq
where we have defined $X_c \equiv |T_c/P_c|, X_t \equiv |T_t/P_t|$.
One consequence of these relations is
\beq \label{eqn:dct}
\cot \delta_t = \cot \delta_c + \frac{\sin \beta}{X_c \sin \alpha \sin
 \delta_c}~~~,
\eeq
which we shall use when relating $\delta_t$ to $\delta_c$.

\section{Measurables in terms of weak and strong phases}

In the present section we derive expressions for the two CP asymmetries
in $\bo(t) \to \pi^+\pi^-$, $\spp$ and $\cpp$, in terms of a strong and a weak 
phase. For completeness, expressions are given in the two equivalent
conventions, which imply identical constraints on $\alpha$.  These
constraints do not require knowledge of the tree/penguin ratio.  Information
about this ratio, which could resolve a certain discrete ambiguity in these 
constraints, can be more useful in one convention than in the other. 
This question is discussed in Section V.

The time-dependent rate of an initially produced $B^0$ decaying to $\pi^+
\pi^-$ at time $t$ is given by \cite{MG}
\beq
\Gamma(B^0(t)\to\pi^+\pi^-) \propto e^{-\Gamma_d t}\left [1 + C_{\pi\pi}
 \cos\Delta(m_d t) - S_{\pi\pi}\sin(\Delta m_d t)\right ]~~.
\eeq
The coefficients of $\sin \Delta m_d t$ and $\cos \Delta m_d t$, measured in
time-dependent CP asymmetries of $\pi^+ \pi^-$ states produced in asymmetric
$e^+ e^-$ collisions at the $\Upsilon(4S)$, are
\beq \label{eqn:CSpipi}
\spp \equiv \frac{2 {\rm Im}(\lpp)}{1 + |\lpp| ^2}~~,~~~
\cpp \equiv \frac{1 - |\lpp|^2}{1 + |\lpp|^2}~~~,
\eeq
where
\beq
\lpp \equiv e^{-2i \beta} \frac{A(\ob \to \pi^+ \pi^-)}
{A(B^0 \to \pi^+ \pi^-)}~~~.
\eeq

The extraction of phases from data on $\spp$ and $\cpp$ now proceeds in
the following manner.  As in Ref.\ \cite{GR02}, we define the charge-averaged
branching ratio,
\beq
\bbpp \equiv
[{\cal B}(B^0 \to \pi^+ \pi^-) + {\cal B}(\ob  \to \pi^+ \pi^-)]/2~~~.
\eeq
We use the convention
\beq\label{B0}
{\cal B}(B^0 \to \pi^+ \pi^-) = |A(\bo \to \pi^+\pi^-)|^2
|\vec p_{\pi\pi}|\tau_0~~~,
\eeq
where $|\vec p_{\pi\pi}|$ is the pion center-of-mass momentum and $\tau_0$ 
is the $\bo$ lifetime.

However, in contrast to the approach of Ref.\ \cite{GR02}, we no longer
normalize this branching ratio with respect to the corresponding tree value,
which is convention-dependent.  Instead, we normalize all amplitudes by the
penguin amplitude $P_c$ or $P_t$, which we have shown to be
convention-independent, up to a CKM factor.

Using broken flavor SU(3) \cite{SU3br} and factorization, the magnitude of the
penguin amplitude is obtained from the $|\Delta S| = 1$ penguin amplitude $P'$
which dominates the decay $B^+ \to K^0 \pi^+$ \cite{GHLR}. That is, our
approach relies on neglecting both rescattering effects in $B^+ \to K^0 \pi^+$
and nonfactorizable contributions in penguin amplitudes.  
Several ways of testing the first assumption were discussed in \cite{rescat}.
We note that this assumption is also made in two detailed theoretical 
schemes for calculating weak hadronic matrix elements \cite{BBNS,KLS}.
In the first scheme \cite{BBNS} factorization of penguin amplitudes is assumed 
to hold to a good approximation and strong phases are small. In the second 
framework \cite{KLS} nonfactorizable terms in penguin amplitudes are strongly 
suppressed, but strong phases are sizable. Thus, while it may seem natural to 
combine the assumption of factorization of penguin amplitudes with small 
strong phases, we will not rely on the latter assumption.  

Within the above assumptions, one obtains for the penguin amplitude $|P_i|~
(i = c, t$) an expression in terms of measurable quantities, 
\beq\label{Pi}
|P_i| = \frac{f_{\pi}}{f_K}\left | \frac{V^*_{ib}V_{id}}{V^*_{ib}V_{is}}
\right | |P'|~~,~~~|P'| = |A(B^+ \to K^0\pi^+)|~~.
\eeq 
Here we use a convention similar to Eq.~(\ref{B0})
\beq\label{B+}
{\cal B}(B^+\to K^0\pi^+) \equiv |A(B^+ \to K^0\pi^+)|^2 
|\vec p_{K\pi}|\tau_+~~~,
\eeq
where $|\vec p_{K\pi}|$ is the $\pi$ or $K$ center-of-mass momentum
and $\tau_+$ is the $B^+$ lifetime.

Applying Eqs.~(\ref{B0}), (\ref{Pi}) and (\ref{B+}), one finds for the
normalized rates \cite{JC2}
\bea\label{eqn:bi}
b_i & \equiv &
\frac{|A(\bo \to \pi^+\pi^-)|^2 + |A(\ob \to \pi^+\pi^-)|^2}{2|P_i|^2}
\nonumber \\
& = & \frac{\bbpp}{{\cal B}(B^+ \to \ko \pi^+)}
\left |\frac{V^*_{ib}V_{is}}{V^*_{ib}V_{id}}\right |^2 \frac{f^2_K}{f^2_{\pi}}
\frac{|\vec p_{K\pi}|}{|\vec p_{\pi\pi}|} \frac{\tau_+}{\tau_0}~~.
\eea
The three measurables, $\spp,~\cpp$ and $\bbpp/{\cal B}(B^+\to K^0\pi^+)$ can 
then be 
expressed in terms of the three parameters $X_i,~\delta_i$ and a weak phase. 
We now display these expressions for the two mentioned conventions. 
\bigskip

\leftline{\bf A.  $c$ convention}
\medskip

In this convention one has
\beq\label{Pc}
|P_c| = \frac{f_\pi}{f_K} \left| \frac{V^*_{cb} V_{cd}}{V^*_{cb} V_{cs}}
\right| |P'| = \frac{f_\pi}{f_K}\frac{\lambda}{1 - \frac{\lambda^2}{2}}
 |A(B^+\to K^0\pi^+)|~~~,
\eeq
where $\lambda = 0.22$ is the parameter describing the hierarchy of CKM
elements \cite{WP}.  Then, noting the
weak and strong phases of $T_c$ and $P_c$, and substituting $\alpha = \pi -
\beta - \gamma$ when convenient, we have
\beq
\lpp = e^{2 i \alpha} \left( \frac{X_c + e^{i \delta_c}
e^{i \gamma}}{X_c + e^{i \delta_c} e^{-i \gamma}} \right)~~~,
\eeq
\beq \label{eqn:bpc}
b_c = X_c^2 + 2 X_c \cos \delta_c \cos \gamma + 1~~~,
\eeq
\beq \label{eqn:sppc}
b_c \spp = X_c^2 \sin 2 \alpha + 2 X_c \cos \delta_c \sin(\beta -
 \alpha) - \sin 2 \beta~~~,
\eeq
\beq \label{eqn:cppc}
b_c \cpp = 2 X_c  \sin \delta_c \sin \gamma~~~.
\eeq
One can use Eq.~(\ref{eqn:bpc}) to eliminate $X_c$ using the experimental
values of $b_c$.  Since $b_c$ is a number significantly greater than 1 [see
Eq.~(\ref{eqn:ratio}) below], only one solution of the quadratic equation is
relevant, and one finds
\beq\label{Xc}
X_c = - \cos \delta_c \cos \gamma + \sqrt{(\cos \delta_c \cos \gamma)^2
+ b_c - 1}~~~.
\eeq
This value can then be substituted into the equations (\ref{eqn:sppc}) and
(\ref{eqn:cppc}) for $\spp$ and $\cpp$ and the resulting values plotted
against one another, e.g., as curves for specific values of $\alpha$
parametrized by $\delta_c$.  We shall exhibit such curves in the next
Section.
\bigskip
 
\leftline{\bf B.  $t$ convention}
\medskip

In the $t$ convention, one has
\beq
|P_t| = \frac{f_\pi}{f_K} \left| \frac{V^*_{tb} V_{td}}{V^*_{tb} V_{ts}}
\right| |P'| = \left| \frac{\sin \gamma}{\sin \alpha} \right| |P_c|
~~~~\Rightarrow ~~~~b_t = b_c\left (\frac{\sin\alpha}{\sin\gamma}
\right )^2~~~,
\eeq
\beq
\lpp = \frac{X_t e^{i \alpha} - e^{i \delta_t}}{X_t e^{-i \alpha}
 - e^{i \delta_t}}~~~,
\eeq
\beq \label{eqn:bpt}
b_t = X_t^2 - 2 X_t \cos \delta_t \cos \alpha + 1~~~,
\eeq
\beq\label{btspp}
b_t \spp = X_t^2 \sin 2 \alpha - 2 X_t \cos \delta_t \sin \alpha ~~~,
\eeq
\beq
b_t \cpp = 2 X_t \sin \delta_t \sin \alpha~~~.
\eeq
In solving Eq.~(\ref{eqn:bpt}) for $X_t$ one again takes the positive
square root:
\beq\label{Xt}
X_t = \cos \delta_t \cos \alpha + \sqrt{(\cos \delta_t \cos \alpha)^2
+ b_t - 1}~~~.
\eeq
Here it is convenient to use the relation $b_t = b_c (\sin \alpha / \sin
\gamma)^2$ since $b_c$ is most directly related to an experimental input. 

Again, one may substitute the value of $X_t$ into the equations for
$\spp$ and $\cpp$ and plot them against one another.  Moreover, in
this convention one may also eliminate both $X_t$ and $\delta_t$, thereby 
obtaining an equation for $\alpha$ alone in terms of measurable quantities:
\bea\label{alphaEq}
b_t \spp &=& \frac{1}{2}\sin 4\alpha + (b_t - 1)
 \sin 2\alpha
\nonumber \\
& \pm & \cos 2\alpha \sqrt{\sin^2 2\alpha + 4(b_t - 1) \sin^2\alpha -
(b_t \cpp)^2}~~.
\eea
This equation is derived in an analogous manner to one obtained recently
for the phase $\gamma$ in terms of measurables in $B_s(t) \to K^+K^-$ and
$B_s \to \ko \ok$ \cite{BsKK}. 
\bigskip

\leftline{\bf C.  Experimental inputs}
\medskip

The most recent measurements of $\spp$ and $\cpp$ \cite{CSBa,bCSBe}, together
with our average of them, are shown in Table \ref{tab:cs}.  
(We have corrected the BaBar entry for $\spp$ misquoted by us in Ref.\
\cite{GR02}.)

\begin{table}
\caption{Values of $\spp$ and $\cpp$ from Refs.\ \cite{CSBa,bCSBe} and their
averages. \label{tab:cs}}
\begin{center}
\begin{tabular}{c c c} \hline \hline
Collab. &          $\spp$           &          $\cpp$           \\ \hline
BaBar   & $-0.01 \pm 0.37 \pm 0.07$ & $-0.02 \pm 0.29 \pm 0.07$ \\
Belle   & $-1.21^{+0.38+0.16}_{-0.27-0.13}$ &
 $-0.94^{+0.31}_{-0.25} \pm 0.09$ \\
Average &      $-0.64 \pm 0.26$     &      $-0.49 \pm 0.21$ \\ \hline \hline
\end{tabular}
\end{center}
\end{table}

The present world averages of $\bbpp$ and ${\cal B}(B^+\to K^0\pi^+)$, 
combining measurements from the CLEO, Belle and BaBar collaborations,
are \cite{RB}
\beq
\bbpp = (5.2 \pm 0.6) \times 10^{-6}~~,~~~~~~~~
{\cal B}(B^+ \to K^0\pi^+) = (17.9 \pm 1.7) \times 10^{-6}~~~.  
\eeq
Adding errors in quadrature, using $f_\pi = 130.7$ MeV, $f_K = 159.8$ MeV and
$\tau_+/\tau_0 = 1.068 \pm 0.016$ \cite{lifetime}, 
we find for the normalized rate in Eq.~(\ref{eqn:bi})
\beq \label{eqn:ratio}
b_c = 9.04 \pm 1.36~~~.
\eeq

\section{CP-violating asymmetries}

\begin{figure}
\centerline{\includegraphics[height=7.5in]{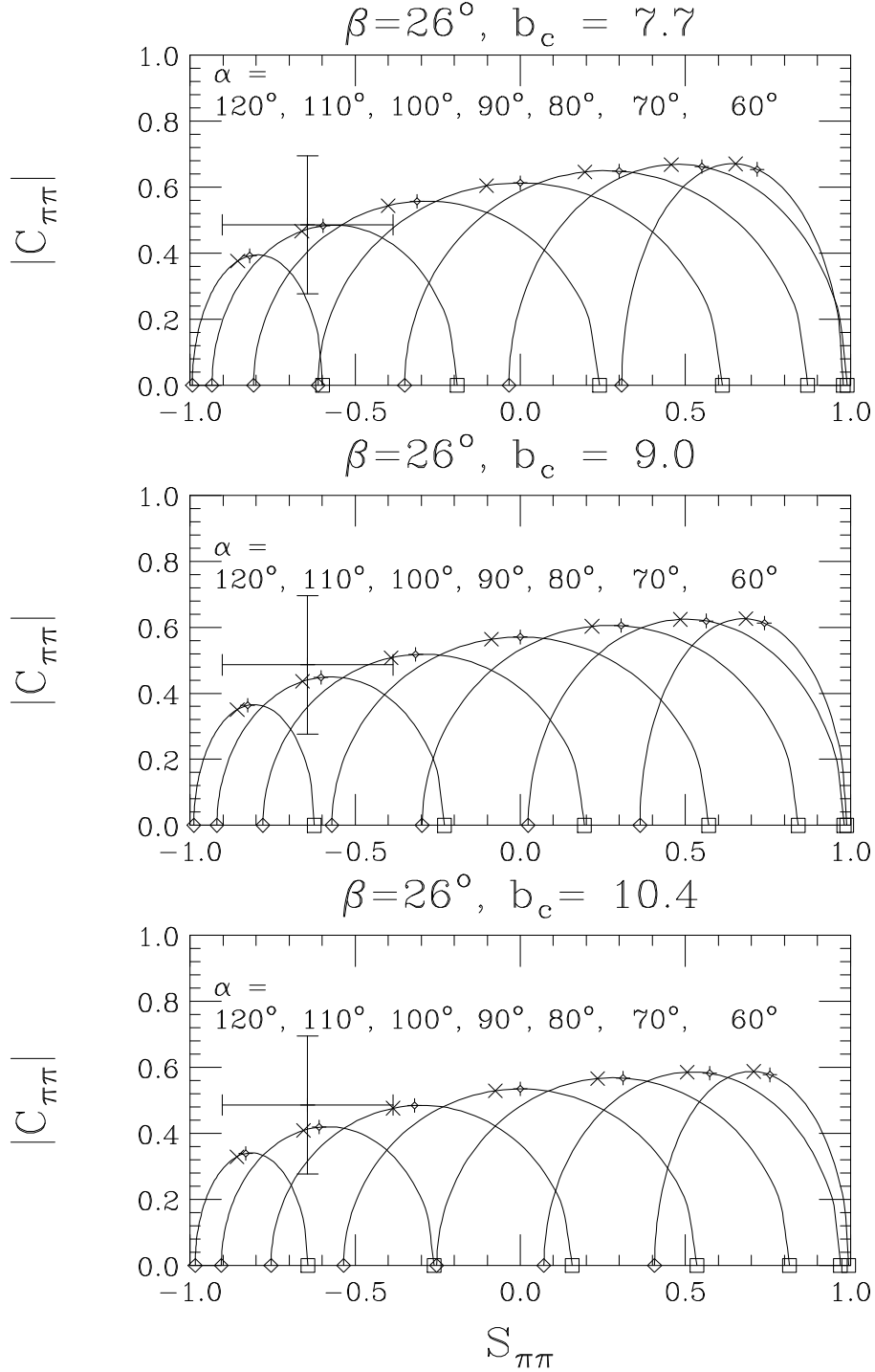}}
\caption{Plots of $|C_{\pi \pi}|$ versus $S_{\pi \pi}$ for various values
of $b_c$.  Top panel:  $b_c = 7.7$.  Middle
panel: $b_c = 9.0$.  Bottom panel: $b_c =
10.4$.  Curves correspond, from left to right, to values of $\alpha$
in $10^\circ$ steps ranging from $120^\circ$ to $60^\circ$.  The value
$\beta = 26^\circ$ has been chosen.  Large plotted point corresponds
to present average of BaBar and Belle data (see text).  Small plotted points:
$\delta_c = \delta_t = 0$ (diamonds), $\delta_c = \delta_t = \pi$ (squares),
$\delta_c = \pi/2$ (crosses), $\delta_t = \pi/2$ (fancy + signs).
\label{fig:cspd}}
\end{figure}

For a given value of $b_c$, Eqs.~(\ref{eqn:sppc})--(\ref{Xc}) [or 
Eqs.~(\ref{btspp})-(\ref{Xt})] can be used to plot $\spp$ and $\cpp$ as 
functions of $\alpha$ and $\delta_c$ (or $\delta_t$).  The values of $\spp$ and
$\cpp$ for the central and $\pm 1 \sigma$ values of the ratio $b_c$ in 
(\ref{eqn:ratio}), and for values of $\alpha$ mostly lying within
the physical range \cite{Beneke} $\alpha = (97^{+30}_{-21})^\circ$, are plotted
in Fig.\ \ref{fig:cspd}.  (For other values of $\alpha$ see, e.g., Ref.\
\cite{GR02}.)  We use $\beta = 26^\circ$ based on the most recent average
$\sin 2 \beta = 0.78 \pm 0.08$ of Belle \cite{bCSBe} and BaBar \cite{betaBa}
values; the $\pm 4^\circ$ error on $\beta$ has little effect \cite{GR02}.
The large plotted point corresponds to the average in Table \ref{tab:cs}.
As expected, the  curves are identical in the two conventions. The existence 
of two 
solutions for $\spp$, for given values of $b_c, \alpha$ and $\cpp$, can be
easily understood. This follows from the $\pm$ sign in Eq.~(\ref{alphaEq}).

For strong phases $\delta_c$ or $\delta_t$ of 0 or $\pi$, the
predictions for $\spp$ and $\cpp$ depend only on $b_c$ and
$\alpha$.  These points are marked with diamonds and squares, respectively.
A strong phase of $\pi$ would signify a relative sign of tree and penguin
amplitudes opposite to that obtained from factorization.
Such a phase is strongly disfavored relative to a zero phase.
For non-zero strong phases, the curves are identical in the two
conventions, but points on them correspond to different values of $\delta_c$
and $\delta_t$.  Examples are shown for $\delta_c = \pi/2$ (crosses) and
$\delta_t = \pi/2$ (fancy $+$ signs).

If $\cpp$ is indeed small, as suggested by the BaBar data \cite{CSBa}, $\alpha$
can be uncertain by as much as about $30^\circ$, depending on whether the
strong phase is near $0$ or $\pi$. This is seen in Fig.~1, where for $b_c=7.7$
the curves for $\alpha = 90^\circ$ and $\alpha = 120^\circ$ intersect near the
horizontal axis. In that case, additional theoretical input \cite{BBNS,KLS}
on strong phases can help resolve the ambiguity.  Theoretically,
it is much more likely that the strong phase is near 0 than near $\pi$.  
If the central value of $\cpp$ remains as large as suggested by the
present experimental average, the discrete ambiguity becomes less of a problem.
Nonetheless, as one can see from neighboring curves,
even a very tiny error ellipse in the $(\spp,\cpp)$ plane will not be able
to resolve values of $\alpha$ differing by $10^\circ$.  This is a necessary
price for giving up prior information on the tree/penguin ratio.

\begin{figure}
\centerline{\includegraphics[height=6.2in]{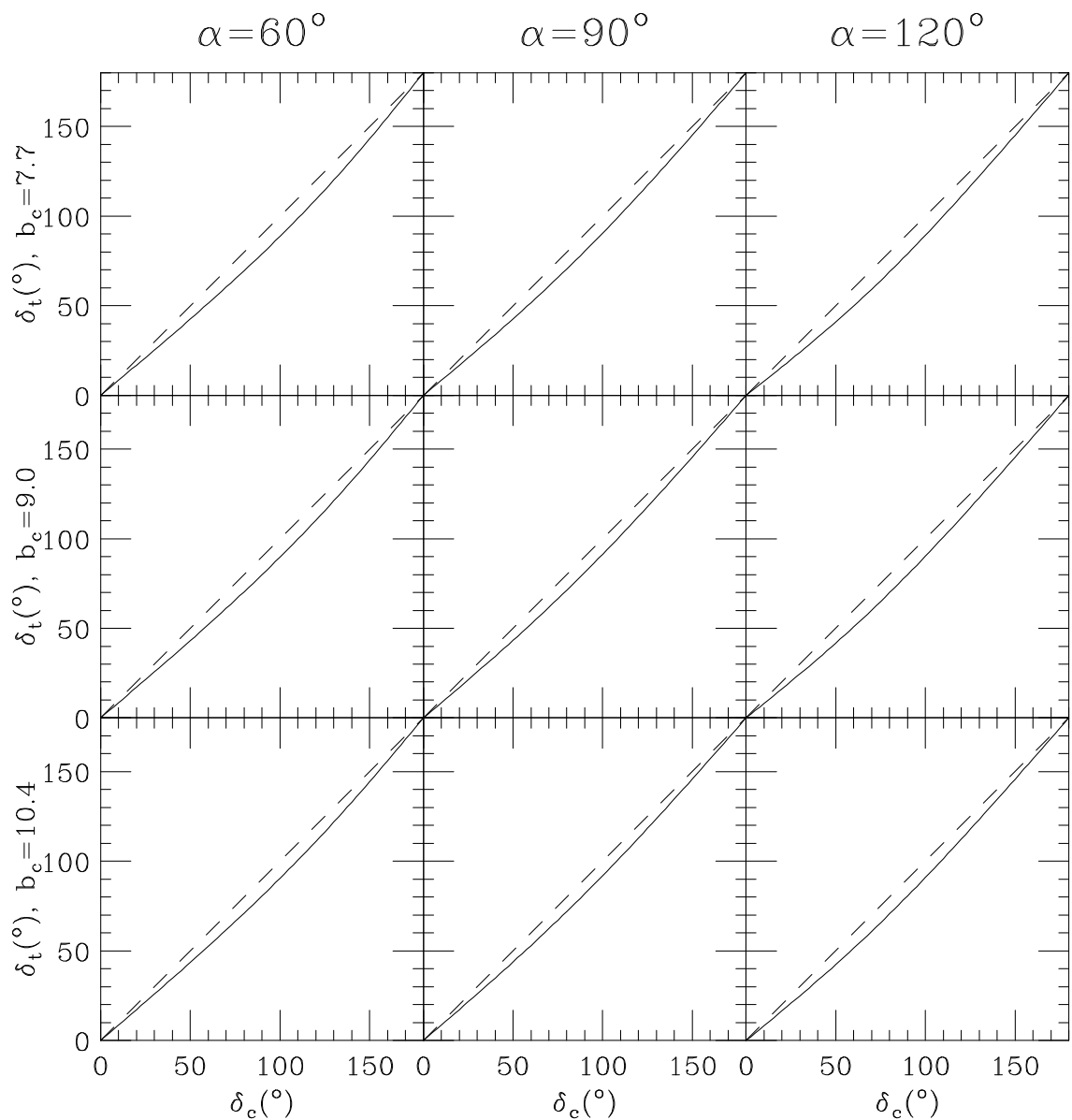}}
\caption{Relations between $\delta_c$ and $\delta_t$ for various values
of $\alpha$ and $b_c$.
\label{fig:ddict}}
\end{figure}

The values of $\delta_c$ and $\delta_t$ do not differ very much from one
another.  When they are close to $\pi/2$, their difference is close to
maximal, but rarely exceeds $10^\circ$, as shown in Fig.\ \ref{fig:ddict}.
We used Eq.~(\ref{eqn:dct}) in making these plots.

We have assumed factorization in obtaining the penguin amplitude. Any deviation
from factorization would result in a corrected value for $b_c$, for which we
have taken a 15\% error arising from experimental errors in branching ratios.
This would be equivalent to correcting the SU(3) breaking factor $f_K/f_\pi$ in
Eq.\ (\ref{eqn:bi}) by 7.5\%.  That is, even assuming perfect measurements of
$\bbpp$ and ${\cal B}(B^+ \to \ko \pi^+)$, an irreducible uncertainty would
be associated with the assumption of factorization for penguin amplitudes.
If this uncertainty were 7.5\%, we would obtain for perfect branching ratio
measurements the range of possibilities shown in Fig.\ \ref{fig:cspd}.

Let us assume that this 7.5\% is a reasonable estimate of the intrinsic
possible deviation from factorization.  By comparing the three panels of Fig.\
\ref{fig:cspd}, one sees that if $\cpp$ is near its maximum, then $\spp$
is not very sensitive to the value of $b_c$ (and hence to the factorization
assumption), while if $\cpp$ is near zero, a given value of $\spp$ corresponds
to values of $\alpha$ differing by only a few degrees depending on the
value of $b_c$ (aside from the much-more-serious discrete ambiguity mentioned
earlier).  In either case, the factorization assumption is not the source
of the limiting error on $\alpha$.

\section{Defining and using a tree/penguin ratio}

Although we have shown that one does not need to know the tree/penguin
ratio in order to extract useful information from $\bbpp$, $\spp$,
and $\cpp$, the error on $\alpha$ and the strong phase $\delta_c$ or
$\delta_t$ can be further reduced if one has some information on 
$X_c$ or $X_t$.  In the present section we first give an example of how
improved information would help, and then discuss the more difficult
questions of which parameter ($X_c$ or $X_t$) is capable of being specified
more precisely and how one would go about doing so.

Let us take as an example an ambiguity associated with curves for $\alpha
= 90^\circ$ and $110^\circ$ which intersect for the central value of
$b_c = 9.0$ around $\spp = -0.4$ and $|\cpp| = 0.4$.
These correspond to different values of $X_c$ or $X_t$, as illustrated in
Table \ref{tab:comp}.  We also show two different values of $\alpha$
($90^\circ$ and $119^\circ$) giving rise to the same values of $\spp$
for $\cpp = 0$.

\begin{table}
\caption{Comparison of $X_c$ and $X_t$ values for pairs of $\alpha$ values
giving the same $\spp$ and $\cpp$.  Here we have taken $b_c = 9.04$.
\label{tab:comp}}
\begin{center}
\begin{tabular}{c c c c c c c} \hline \hline
$\alpha$   & $\spp$ & $|\cpp|$ & $X_c$ & $\delta_c$ & $X_t$ & $\delta_t$ \\
 \hline
$90^\circ$ & $-0.41$ &  0.40  & 2.6 & $51^\circ$  & 3.2 & $44^\circ$  \\
$110^\circ$ & $-0.41$ & 0.40  & 3.3 & $129^\circ$ & 4.1 & $122^\circ$ \\
 \hline
$90^\circ$ & $-0.57$ & 0.0 & 2.4 & $0^\circ$ & 3.2 & $0^\circ$ \\
$119^\circ$ & $-0.57$ & 0.0 & 3.8 & $180^\circ$ & 4.9 & $180^\circ$ \\
 \hline \hline
\end{tabular}
\end{center}
\end{table}

From these examples, one sees that specification of $X_c$ or $X_t$ with an
error of $\pm 0.3$ would permit resolution of the ambiguity.  In Ref.
\cite{GR02} we employed an estimate $X_c \simeq 3.6$ with about a 25\%
error.  Reduction of this error to about $\pm 10\%$ is needed in order to
have a significant impact on resolving the ambiguity exhibited in
Table \ref{tab:comp}.  Is such accuracy achievable?

Our estimate of $b_c$ involves a 15\% error which consists of slightly less
than 10\% due to that in ${\cal B}(B^+ \to K^0 \pi^+)$, and slightly more than
10\% due to that in $\bbpp$, added in quadrature.  Clearly these errors will
shrink with improved statistics.  However, the determination of $|T_c|$
from $B \to \pi \ell \nu$ using factorization is problematic 
since $T_c \sim A_u - A_t$ [Eq.\ (4)] contains the short-distance
penguin contribution involving the top quark loop.  It might seem more
reliable to estimate $T_t \sim A_u - A_c$ [Eq.\ (5)] using factorization 
since its penguin contribution does not contain a large logarithm of $m_t$.
This is in fact the method advocated in Ref.\ \cite{LRpipi}, in which a
determination of $T_t$ with an accuracy of less than 6\% was deemed feasible
with about 500 $B \to \pi l \nu$ events.  A corresponding accuracy for
$|P_t|$ would require improved accuracy for ${\cal B}(B^+ \to K^0 \pi^+)$
(which gives $|P_c|$, not $|P_t|$) and then using the relation (\ref{PtPc}), 
$|P_t| = |P_c| \sin \gamma/ \sin \alpha$.

A potential problem with determining $T_t$ using
factorization is that while its contamination from the short-distance
penguin amplitude is less than that in $T_c$, there is no corresponding
guarantee for {\it long-distance} penguin contributions such as might be
introduced by rescattering from tree amplitudes, for example
via $B^0 \to D^{(*)+}  D^{(*)-} \to \pi^+ \pi^-$.  Other processes, such as
$B^0 \to K^+ K^-$, are expected to proceed {\it mainly} via rescattering or
else, if rescattering is unimportant, to be highly suppressed \cite{rescat}.
Present bounds on this last process are quite stringent
\cite{KKlim}:
${\cal B}(B^0 \to K^+ K^-) \le 0.5 \times 10^{-6}$.  It may be that one
must rely on theoretical treatments of factorization (e.g., \cite{BBNS})
in order to specify $|T_t|$ (or perhaps $|T_c|$) more precisely. 

\section{Experimental prospects and conclusions}

We have shown that one can obtain useful information on weak and strong
phases by studying the observables in $B^0(t) \to \pi^+ \pi^-$ without
having to define in advance the ratio of tree and penguin amplitudes,
and in a manner which is independent of the convention adopted for the
penguin amplitudes.  These observables consist of the flavor-averaged
branching ratio $\bbpp$ normalized by ${\cal B}(B^+ \to \ko \pi^+)$
and the quantities $\spp$ and $\cpp$ measured in
time-dependent asymmetries.  We consider only information based on the
magnitude of $\cpp$; its sign determines the sign of the strong phase shift.

The degree of information obtainable without auxiliary tree/penguin
information can be estimated from the curves in Figure 1 and depends on
whether $|\cpp|$ is near its maximum value (the envelope of the curves)
or zero.  If $|\cpp| \simeq 0$, important discrete ambiguities in $\alpha$
exist, amounting to up to about $30^\circ$, which must be resolved using
additional information on the tree/penguin ratio or on the strong phase.  If
$|\cpp|$ is near its maximum, the error on $\alpha$ appears to depend
roughly on the square root of the error in $|\cpp|$, as one can see by
measuring how far from the envelope of the curves the intersection point
of two curves for different $\alpha$ values lies.  Thus, two curves for
$\alpha$ differing by $(10,20,30)^\circ$ intersect at points about $(0.04,%
0.08,0.18)$ below the envelope along the $|\cpp|$ axis.  To take one example,
if one wants to distinguish between two curves for $\alpha$ differing by
$20^\circ$ (as in the example of Table \ref{tab:comp}), one should be
prepared to measure $|\cpp|$ with an error of no more than $\pm 0.08$, which
is about 2.6 times less than the present error of $\pm 0.21$.  One thus
would need $(2.6)^2$ times the data sample ($\simeq 100$ fb$^{-1}$) on which
Table \ref{tab:cs} was based, or about 700 fb$^{-1}$ from the total of
BaBar and Belle.  This appears to be within the goals of the experiments.
Errors on $\spp$ in such a sample should be sufficiently small that they
will not play a major role in the errors in $\alpha$.

\section*{Acknowledgments}

M. G. wishes to thank The Enrico Fermi Institute at the University of Chicago
for its kind hospitality.  We thank A. H\"ocker for asking the question which
led to this investigation, H. Jawahery for a communication regarding data,
and H. N. Li, S. Olsen and L. Wolfenstein for helpful discussions.
This work was supported in part by the United States Department of Energy
through Grant No.\ DE FG02 90ER40560, by the Israel Science Foundation founded
by the Israel Academy of Sciences and Humanities, and by the US - Israel
Binational Science Foundation through Grant No. 98-00237.

\def \ajp#1#2#3{Am.\ J. Phys.\ {\bf#1}, #2 (#3)}
\def \apny#1#2#3{Ann.\ Phys.\ (N.Y.) {\bf#1}, #2 (#3)}
\def \app#1#2#3{Acta Phys.\ Polonica {\bf#1}, #2 (#3)}
\def \arnps#1#2#3{Ann.\ Rev.\ Nucl.\ Part.\ Sci.\ {\bf#1}, #2 (#3)}
\def \art{and references therein}
\def \cmts#1#2#3{Comments on Nucl.\ Part.\ Phys.\ {\bf#1}, #2 (#3)}
\def \cn{Collaboration}
\def \cp89{{\it CP Violation,} edited by C. Jarlskog (World Scientific,
Singapore, 1989)}
\def \econf#1#2#3{Electronic Conference Proceedings {\bf#1}, #2 (#3)}
\def \efi{Enrico Fermi Institute Report No.}
\def \epjc#1#2#3{Eur.\ Phys.\ J.\ C {\bf#1}, #2 (#3)}
\def \f79{{\it Proceedings of the 1979 International Symposium on Lepton and
Photon Interactions at High Energies,} Fermilab, August 23-29, 1979, ed. by
T. B. W. Kirk and H. D. I. Abarbanel (Fermi National Accelerator Laboratory,
Batavia, IL, 1979}
\def \hb87{{\it Proceeding of the 1987 International Symposium on Lepton and
Photon Interactions at High Energies,} Hamburg, 1987, ed. by W. Bartel
and R. R\"uckl (Nucl.\ Phys.\ B, Proc.\ Suppl., vol. 3) (North-Holland,
Amsterdam, 1988)}
\def \ib{{\it ibid.}~}
\def \ibj#1#2#3{~{\bf#1}, #2 (#3)}
\def \ichep72{{\it Proceedings of the XVI International Conference on High
Energy Physics}, Chicago and Batavia, Illinois, Sept. 6 -- 13, 1972,
edited by J. D. Jackson, A. Roberts, and R. Donaldson (Fermilab, Batavia,
IL, 1972)}
\def \ijmpa#1#2#3{Int.\ J.\ Mod.\ Phys.\ A {\bf#1}, #2 (#3)}
\def \ite{{\it et al.}}
\def \jhep#1#2#3{JHEP {\bf#1}, #2 (#3)}
\def \jpb#1#2#3{J.\ Phys.\ B {\bf#1}, #2 (#3)}
\def \lg{{\it Proceedings of the XIXth International Symposium on
Lepton and Photon Interactions,} Stanford, California, August 9--14, 1999,
edited by J. Jaros and M. Peskin (World Scientific, Singapore, 2000)}
\def \lkl87{{\it Selected Topics in Electroweak Interactions} (Proceedings of
the Second Lake Louise Institute on New Frontiers in Particle Physics, 15 --
21 February, 1987), edited by J. M. Cameron \ite~(World Scientific, Singapore,
1987)}
\def \kaon{{\it Kaon Physics}, edited by J. L. Rosner and B. Winstein,
University of Chicago Press, 2001}
\def \kdvs#1#2#3{{Kong.\ Danske Vid.\ Selsk., Matt-fys.\ Medd.} {\bf #1}, No.\
#2 (#3)}
\def \ky{{\it Proceedings of the International Symposium on Lepton and
Photon Interactions at High Energy,} Kyoto, Aug.~19-24, 1985, edited by M.
Konuma and K. Takahashi (Kyoto Univ., Kyoto, 1985)}
\def \mpla#1#2#3{Mod.\ Phys.\ Lett.\ A {\bf#1}, #2 (#3)}
\def \nat#1#2#3{Nature {\bf#1}, #2 (#3)}
\def \nc#1#2#3{Nuovo Cim.\ {\bf#1}, #2 (#3)}
\def \nima#1#2#3{Nucl.\ Instr.\ Meth.\ A {\bf#1}, #2 (#3)}
\def \npb#1#2#3{Nucl.\ Phys.\ B~{\bf#1}, #2 (#3)}
\def \npps#1#2#3{Nucl.\ Phys.\ Proc.\ Suppl.\ {\bf#1}, #2 (#3)}
\def \os{XXX International Conference on High Energy Physics, Osaka, Japan,
July 27 -- August 2, 2000}
\def \PDG{Particle Data Group, D. E. Groom \ite, \epjc{15}{1}{2000}}
\def \pisma#1#2#3#4{Pis'ma Zh.\ Eksp.\ Teor.\ Fiz.\ {\bf#1}, #2 (#3) [JETP
Lett.\ {\bf#1}, #4 (#3)]}
\def \pl#1#2#3{Phys.\ Lett.\ {\bf#1}, #2 (#3)}
\def \pla#1#2#3{Phys.\ Lett.\ A {\bf#1}, #2 (#3)}
\def \plb#1#2#3{Phys.\ Lett.\ B {\bf#1}, #2 (#3)}
\def \prl#1#2#3{Phys.\ Rev.\ Lett.\ {\bf#1}, #2 (#3)}
\def \prd#1#2#3{Phys.\ Rev.\ D\ {\bf#1}, #2 (#3)}
\def \prp#1#2#3{Phys.\ Rep.\ {\bf#1}, #2 (#3)}
\def \ptp#1#2#3{Prog.\ Theor.\ Phys.\ {\bf#1}, #2 (#3)}
\def \rmp#1#2#3{Rev.\ Mod.\ Phys.\ {\bf#1}, #2 (#3)}
\def \rp#1{~~~~~\ldots\ldots{\rm rp~}{#1}~~~~~}
\def \si90{25th International Conference on High Energy Physics, Singapore,
Aug. 2-8, 1990}
\def \slc87{{\it Proceedings of the Salt Lake City Meeting} (Division of
Particles and Fields, American Physical Society, Salt Lake City, Utah, 1987),
ed. by C. DeTar and J. S. Ball (World Scientific, Singapore, 1987)}
\def \slac89{{\it Proceedings of the XIVth International Symposium on
Lepton and Photon Interactions,} Stanford, California, 1989, edited by M.
Riordan (World Scientific, Singapore, 1990)}
\def \smass82{{\it Proceedings of the 1982 DPF Summer Study on Elementary
Particle Physics and Future Facilities}, Snowmass, Colorado, edited by R.
Donaldson, R. Gustafson, and F. Paige (World Scientific, Singapore, 1982)}
\def \smass90{{\it Research Directions for the Decade} (Proceedings of the
1990 Summer Study on High Energy Physics, June 25--July 13, Snowmass,
Colorado),
edited by E. L. Berger (World Scientific, Singapore, 1992)}
\def \tasi{{\it Testing the Standard Model} (Proceedings of the 1990
Theoretical Advanced Study Institute in Elementary Particle Physics, Boulder,
Colorado, 3--27 June, 1990), edited by M. Cveti\v{c} and P. Langacker
(World Scientific, Singapore, 1991)}
\def \TASI{{\it TASI-2000:  Flavor Physics for the Millennium}, edited by J. L.
Rosner (World Scientific, 2001)}
\def \yaf#1#2#3#4{Yad.\ Fiz.\ {\bf#1}, #2 (#3) [Sov.\ J.\ Nucl.\ Phys.\
{\bf #1}, #4 (#3)]}
\def \zhetf#1#2#3#4#5#6{Zh.\ Eksp.\ Teor.\ Fiz.\ {\bf #1}, #2 (#3) [Sov.\
Phys.\ - JETP {\bf #4}, #5 (#6)]}
\def \zpc#1#2#3{Zeit.\ Phys.\ C {\bf#1}, #2 (#3)}
\def \zpd#1#2#3{Zeit.\ Phys.\ D {\bf#1}, #2 (#3)}


\begin{thebibliography}{99}
\bibitem{JC2} A ratio similar to $b_c$, denoted by Charles \cite{PT} as 
$1/R_P$, did not contain the last three factors in the second line and 
assumed $|V_{cs}|^2 =1$.
\end{thebibliography}
\end{document}